\documentstyle[eqsecnum,aps,prd,epsfig,cite]{revtex}
\date{\today}
\def\be{\begin{equation}}
\def\bea{\begin{eqnarray}}
\def\eea{\end{eqnarray}}
\def\ee{\end{equation}}
\def\no{\nonumber}
\def \bU{{\bf U}}
\def \bV{{\bf V}}
\def \bC{{\bf C}}
\def \bp{{\bf p}}
\def \bq{{\bf q}}
\def \bmu{{\bf \mu}}
\def \bA{{\bf A}}
\def \D{\Delta}
\def \un{{\bf{\hat n}}}
\def \tC{{\tilde C}}
\def \a {\alpha}
\def \b {\beta}
\def \g {\gamma}
\def \z {\zeta}
\def \A {{\cal A}}
\def \cU{{\cal U}}

\begin{document}
\title{\bf {Algebraic approach to time-delay data analysis for LISA}}
\author{S. V. Dhurandhar$^1$, K. Rajesh Nayak$^1$ and J-Y. Vinet$^2$}
 
\address{ $^1$IUCAA, Postbag 4, Ganeshkind, Pune - 411 007, India. 
\\ $^2$ILGA, Dept. Fresnel, Observatoire de la Cote d'Azur, 
\\ BP 4229, 06304 Nice, France.
}

\maketitle

\begin{abstract}
Cancellation of laser frequency noise in interferometers is crucial for attaining the 
requisite sensitivity of the triangular 3-spacecraft LISA configuration. Raw laser noise 
is several orders of magnitude above the other noises and thus it is essential to bring it
down to the level of other noises such as shot, acceleration, etc. Since it is 
impossible to maintain equal distances between spacecrafts, laser noise 
cancellation must be achieved by appropriately combining the six beams with appropriate 
time-delays.  It has been shown in several recent
papers that such combinations are possible. In this paper, we present a rigorous and systematic formalism based on 
algebraic geometrical methods involving computational commutative algebra, which generates 
in principle {\it all} the data combinations cancelling the laser frequency noise. The 
relevant data combinations form the first module of syzygies, as it is called in the 
literature of algebraic geometry. The module is over a polynomial ring in three variables, 
the three variables corresponding to the three time-delays around the LISA triangle.
Specifically, we list several sets of generators for 
the module whose linear combinations with polynomial coefficients generate the entire 
module. We find that this formalism can also be extended in a straight forward way to cancel 
Doppler shifts due to optical bench motions. The two modules are infact isomorphic.
\par
We use our formalism to obtain the transfer functions for the six beams and for the 
generators. We specifically investigate monochromatic gravitational wave sources in the 
LISA band and carry out the maximisiation over linear combinations 
of the generators
of the signal-to-noise ratios with the frequency and source direction angles as 
parameters.
\end{abstract}

\section{Introduction}

  Breakthroughs in modern technology have made  possible the construction of  
extremely large interferometers both on ground and in space for the detection 
and observation of gravitational waves (GW). Several ground based detectors are 
being constructed around the globe; these are the projects,  
LIGO, VIRGO, GEO, TAMA and AIGO of building interferometers whose 
armlengths will be of the order of kilometers. These detectors will operate in the 
high frequency range of GW of $\sim 10$ Hz to a few kHz. A natural limit occurs on 
decreasing the lower frequency cut-off of $10$ Hz because it is not practical to 
increase the armlengths on ground and also because of the gravity gradient noise 
which is difficult to eliminate below $10$ Hz. Thus, the ground based interferometers 
will not be sensitive below the limiting frequency of $\sim 10$ Hz.  But on the otherhand, 
there exist in the cosmos, interesting astrophysical GW sources which emit GW below this 
frequency such as the galactic binaries, massive and supermassive blackhole binaries etc.
If we wish to observe these sources, we need to go to lower frequencies. 
The solution is to build an interferometer in space, where such noises will be absent and 
allow the detection of GW in the low frequency regime. LISA - Laser Interferometric Space 
Antenna - is a proposed mission which will use coherent laser beams exchanged between three 
identical spacecrafts forming a giant (almost) equilateral triangle of side 
$5 \times 10^6$ kilometers to observe and detect low 
frequency cosmic GW. The ground based detectors and LISA complement each other in the 
observation of GW in an essential way, analogous to the optical, radio, X-ray, 
$\gamma$-ray etc., observations do for the electromagnetic waves. As these detectors begin 
to operate, a new era of {\it Gravitational Astronomy} is on the horizon and a radically 
different view of the universe is expected to be revealed.  
\par
In ground based detectors the arms are chosen to be of equal length so that the laser light 
experiences identical delay in each arm of the interferometer. This arrangement precisely 
cancels the laser frequency/phase noise at the photodetector. This cancellation of noise is 
crucial since the raw laser noise is orders of magnitude larger than other noises in the 
interferometer. The required sensitivity of the instrument can thus only be achieved by 
near exact cancellation of the laser frequency noise. However, in LISA it is impossible to 
achieve equal distances between spacecrafts and the laser noise cannot be cancelled in an 
obvious manner. In LISA, six data streams arise from the exchange of laser beams between 
the three spacecrafts - it is not possible to bounce laser beams between different 
spacecrafts, as is done in ground based detectors; a scheme analogous to the RF transponder 
scheme is used as was done in the early experiments for detecting GW by Doppler tracking 
a spacecraft from Earth. Several schemes, some quite elaborate, have been proposed 
\cite{AET99,ETA00} which combine the recorded data by suitable time-delays 
corresponding to the 
three armlengths of the giant triangular interferometer. These schemes  have 
data combinations of six or at most eight data points which give respectively a six and 
eight pulse response of GW and also show how other data combinations can be obtained by 
linear superposition. 
\par
Galactic and extragalactic binaries  are important sources  in the LISA
frequency band.  Their  abundance  and resulting spectral amplitude  has
been  estimated using the population synthesis by various 
authors[\citen{PBDH}-\citen{NYPZ}]. In the lower frequency range 
($ \leq 1$ mHz) there are a large number of such  sources  in each of the  
frequency bins.  This makes it impossible to resolve  an  individual  
source which results in a  stochastic GW background. 
It has been also proposed  that the galactic halo MACHOs such as white 
dwarfs and blackholes ( with mass $ \sim 0.5~M_{\odot}~$) can also 
produce stochastic GW background[\citen{WH}-\citen{HSR}].  In a recent
work, 
Tinto {\it at. al}\cite{TAE} have used Doppler delayed beams for  
discriminating the stochastic  background  from the instrumental noise.
The angular resolution of  the
LISA  is restricted because it is an all-sky monitoring detector with 
quadrupole beam pattern, however, the angular resolution can be achieved
by the relative amplitude and phases of the two polarisations and Doppler
modulation of the beams due to the  motion of LISA around the 
sun\cite{CUTL,RSCH}. 
\par
We start with the fundamental papers by the JPL team \cite{AET99,ETA00,TA99} where idea 
of delayed data combinations was first proposed. 
Here we present a {\it systematic method} based on modules over polynomial rings, 
which not only reproduces the previously obtained results, but guarantees {\it all} 
the data combinations which cancel the laser noise. The data combinations in the 
case of laser frequency noise consist 
of the six suitably delayed data streams, the delays being integer multiples of the 
light travel times between spacecrafts, which can be conveniently expressed in terms 
of polynomials in the three delay operators $E_1, E_2, E_3$ corresponding to the light 
travel time along the three arms. The laser noise cancellation condition puts three 
constraints on the six 
polynomials of the delay operators corresponding to the six data streams. The  
problem therefore consists of finding six tuples of polynomials which satisfy 
the laser noise cancellation constraints. These polynomial tuples form a 
module\footnote{A module is an abelian group over a {\it ring} as contrasted with a vector 
space which is an abelian group over a field. The scalars form a ring and just like in a 
vector space, scalar multiplication is defined. However, in a ring the multiplicative 
inverses do not exist in general for the elements, which makes all the difference!} 
called in the literature, the {\it module of syzygies}. There exist standard methods 
for obtaining the module, by that we mean, methods for obtaining the generators of the 
module so that the linear combinations of the generators generate the entire module. 
Three constraints on six tuples of polynomials do not lead to three generators as naive 
reasoning might suggest. Here we are dealing with modules rather than 
vector spaces and the rules are different. The procedure first consists of obtaining 
a Groebner basis for the ideal generated by the coefficients appearing in the 
constraints. This ideal is in the polynomial ring in $E_1, E_2, E_3$ over the 
domain of rational numbers (or integers if one gets rid of the denominators). To 
obtain the Groebner basis for the ideal, one may use the Buchberger algorithm or use 
a package such as Mathematica. From the Groebner basis there is a standard way to obtain   
a generating set for the required module. All of this procedure has been described 
in the literature \cite{becker,martin}. We thus obtain seven generators for the module. 
However, the method does not guarantee a minimal set and we find that a generating set of 
4 polynomial six tuples suffice to generate the required module. Alternatively, we 
can obtain generating sets by using the software Macaulay 2. It gives us a 
a Groebner basis for the module consisting of five generators and another 
generating set consisting of six elements. The importance 
of obtaining more data combinations is evident: they provide the necessary redundancy - 
different data combinations produce 
different transfer functions for GW and so specific data combinations could be optimal 
for given astrophysical source parameters in the context of maximising signal-to-noise 
(SNR), detection probability, improving parameter estimates etc. 
\par
The scheme we have described above can also be extended in a straight forward way to 
include optical bench motions. There are now twelve Doppler streams of data and we 
apply the above scheme to cancel the noise due to optical bench drift and laser 
frequency noise. The six extra streams can be combined two by two by subtracting 
one stream from the other to obtain three streams in which the frequency shifts in 
the optical fibers are cancelled. Thus we have only nine streams to contend with and 
now the module consists of nine tuples of polynomials on which six linear 
constraints are imposed. We show that the problem can be solved in the terms of the 
previous one where the three extra polynomials are written in terms of the six tuple  
polynomials which are solutions to the laser frequency noise cancellation problem. 
Thus the solution to the first problem extends easily to the second.
\par
Finally, we apply our formalism to a class of important astrophysical sources, but 
relatively simple to analyse, namely, monochromatic GW sources. We maximise the SNR 
for such sources over much of the module of data combinations by considering linear 
combinations of the generators of the module with the coefficients being real numbers. 
Strictly speaking one must take polynomials as the coefficients so that the maximisation 
extends to the entire module, but we find that even this simplifying assumption yields 
satisfactory results. We present the maximised SNR as a function of 
frequency over the data combinations.

We  organise the paper in the following manner: In the section II,  we present
the six raw  data  streams obtained with the laser phase noise  and  formulate
the conditions for the laser phase noise cancellation. We also obtain
difference equations  which should be satisfied by  the time-delay
operators  for cancelling the laser noise. 
 The solutions for the noise cancellation combinations
can be represented as the modules syzygies over the polynomial ring using
standard methods of algebraic geometry  described in the section III. 
First a  Groebner basis for the ideal is obtained. From the Groebner basis the 
generators for the module of syzygies can be computed.
 Several sets of generators have
been  obtained  for this module. In the subsection III B this approach is
extended to cancel the acceleration noise of the optical benches. In the
section IV we compute the detector response for the GW signal and obtain
transfer functions for the six elementary beams. In  the section V,
first, we determine the effective noise for each  generator by taking shotnoise
and acceleration noise of the proof masses into account, which do not  cancel 
out in the combinations. We obtain SNRs  for monochromatic sources and
maximise the SNR over the allowed data combinations that  cancel
the laser frequency noise.

\section{Time Delayed Data and the Difference Equation}
%
\subsection{Time-delayed Data}
 We label the spacecrafts as 1,2 and 3. Let $L_1, L_2, L_3$ be the lengths of the arms 
(sides of the triangle) where $L_3$ is the distance between spacecrafts 1 and 2; and so on 
by cyclic rotation of indices (see figure \ref{fig1}). Each spacecraft has a laser which is

\begin{itemize}
 
\item  used to send beams to the other two spacecrafts, and,  

\item used as a local oscillator to produce a beat signal with the incoming beams 
from the other two spacecrafts.

\end{itemize}

The data are recorded as {\it fractional Doppler shifts}. These fractional Doppler shifts 
can occur due to the GW signal and the noise. Here we will be concerned with the laser 
frequency noise only. More precisely, if $\nu_0$ is the central frequency of the laser 
and the 
frequency fluctuation of the laser on spacecraft $i$ at time $t$ is $\Delta \nu_i (t)$, 
then the fractional frequency fluctuation $C_i (t)$ is given by,
\be
C_i (t) = {\Delta \nu_i (t) \over \nu_0}. 
\ee
The six streams of Doppler data are obtained from the $C_i (t)$ by combining them suitably 
with their time delayed copies, where the time delays are just the light travel times 
between the three spacecrafts. We adopt the following notation for the six streams: we 
divide the six streams into two sets $U^i$ and $V^i$ where, $i = 1,2,3$ of three streams 
each. $U^i$ and $V^i$ can be regarded as 3 component vectors $\bU$ and $\bV$ respectively. 
Also we denote the 
time-delayed data in arm $k$, $k = 1,2,3,$ by the shift operator $E_k$ whose action on a 
function $f(t)$ is described by the equation:
\begin{figure}[b]
\begin{center}
\epsfig{file=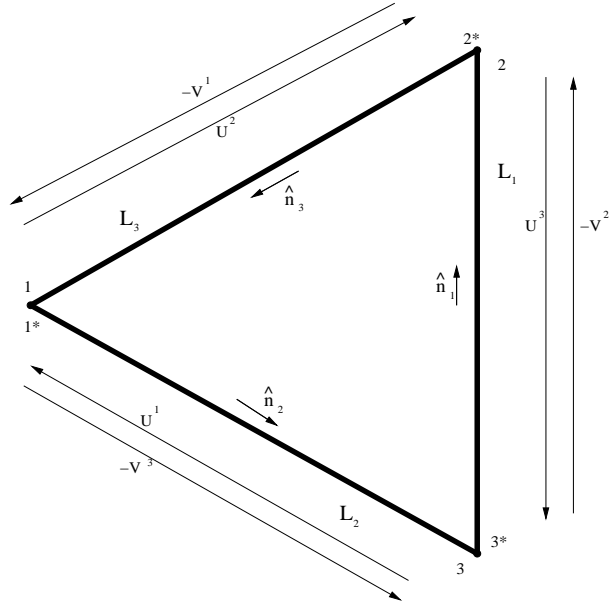,height=8cm,width=8cm}
\caption{LISA Geometry} \label{fig1}
\end{center}
\end{figure}
\be
E_k f(t) = f(t - L_k),
\ee
where we have chosen units in which the speed of light is unity. The six streams are: 
\bea
U^1 &=& E_2 C_3 - C_1, \no \\
U^2 &=& E_3 C_1 - C_2,  \no \\
U^3 &=& E_1 C_2 - C_3,  \no \\
V^1 &=& C_1 - E_3 C_2,  \no \\
V^2 &=& C_2 - E_1 C_3, \no \\
V^3 &=& C_3 - E_2 C_1.
\label{beams} 
\eea  
Thus explicitly we have, $U^1 (t) = C_3 (t - L_2) - C_1 (t)$ etc. $U^1 (t)$ is the 
data stream obtained by beating the laser beam transmitted by spacecraft 3 to spacecraft 1 
measured at time $t$ at spacecraft 1; and so on by cyclic rotation. 
Similarly $- V^1 (t) = C_2 (t - L_3) - C_1 (t)$ 
is the laser beam emitted by spacecraft 2 and received and beaten with the laser at 
spacecraft 1 at time $t$. If we label the spacecrafts in a counter-clockwise (positive 
sense) fashion, then the beams $U^i$ travel in the positive sense while the beams $V^i$ 
travel in the negative sense. 
\par
{\it The goal of the analysis is to add suitably delayed beams together so that 
the net result is zero.} This amounts to seeking data combinations which cancel the 
laser frequency noise. In the notation/formalism that we have invoked, 
the delay is obtained by applying the operators $E_k$ to the beams $U^i$ and $V^i$. A  
delay of $k_1 L_1 + k_2 L_2 + k_3 L_3$ is represented by the operator 
$E_1^{k_1} E_2^{k_2} E_3^{k_3}$ acting on the data, where $k_1, k_2$ and $k_3$ are integers. 
In general a polynomial in $E_k$, which is a polynomial in 
three variables, applied to say $U^1$ combines the same data stream $U^1 (t)$ with different 
time-delays of the form  $k_1 L_1 + k_2 L_2 + k_3 L_3$. This notation 
conveniently rephrases the problem. One must find six 
polynomials say $p_i (E_1, E_2, E_3), \  q_i (E_1, E_2, E_3), \ i=1,2,3 $ 
such that: 
\be
\sum_{i = 1}^3 p_i V^i + q_i U^i = 0.
\label{cnl}
\ee     
Cancellation of the laser noise is implicit in the above equation.

\subsection{The Difference Equation for Shift Operators}

  It is convenient to express eq. (\ref{beams}) in matrix form. This allows us to 
obtain a matrix operator equation whose solutions are $\bp$ and $\bq$ where we have now 
written $p^i$ and $q^i$ as column vectors. We can similarly express $U^i, V^i, C^i$ as 
column vectors $\bU, \bV, \bC$ respectively. In matrix form eq. (\ref{beams}) become:
\be 
\bV = \bmu^T \  \bC, ~~~~~~~ \bU = - \bmu \  \bC,
\ee
where, $\bmu$ is a $3 \times 3$ matrix given by,
\be
\bmu = \left(\begin{array}{ccc}
  1 & 0 & -E_2 \\
  -E_3 & 1 & 0 \\
  0 & -E_1 & 1
\end{array}\right) \,.
\label{mumat}
\ee
The exponent `T' represents the transpose of the matrix. The eq. (\ref{cnl}) becomes:
\be
(\bmu \bp - \bmu^T \bq )^T \ \bC = 0  \, .
\ee
where we have taken care to put $\bC$ on the right of the operators. Since the above 
equation must be satisfied 
for arbitrary `data' $\bC$, we obtain a matrix equation for the shift operators:
\be
\bmu \bp = \bmu^T \bq  \, .
\label{opeq}
\ee
The solutions to eq. (\ref{opeq}) are $\bp, \bq$ which are column vectors of polynomials 
in the shift operators $E_k$. Note that since the $E_k$ are just shift operators, they 
commute, so the order in writing these operators is unimportant. In mathematical terms, 
the polynomials form a commutative ring. 
\par 
We can formally solve for $\bp$  since the matrix $\bmu$ 
is invertible. However, $\det \bmu = 1 - E_1 E_2 E_3$ appears in the denominator on the 
R.H.S., which leads to the division by polynomials in $E_k$. This may seem hard to 
interpret. But we can pull this factor to the L. H. S. to `rationalise' the expressions. 
Then we obtain,
\be
\D \bp = \bA \ \bq \, ,
\label{simpeq}
\ee
where $\bA = \bmu_{adj} \bmu^T$ and $\bmu_{adj}$ is the adjoint of $\bmu$. The operator 
$\D = 1 - E_1 E_2 E_3$ is the usual difference operator that appears in finite 
differences and difference equations. The quantity $E_1 E_2 E_3$ plays a 
central role in determining the natural time-step for the problem, namely, 
$s = L_1 + L_2 + L_3$; which is nothing but the light travel time around the perimeter 
of the LISA triangle. $\D$ is just the difference corresponding to this time-step.
\par 
Explicitly, using (\ref{mumat}) the matrix $\bA$ is given by:
\be
\bA = \left(\begin{array}{ccc} 
1 - E_2^2 & E_1 E_2 - E_3 & E_2 (1 - E_1^2) \\
(1 - E_2^2) E_3 & 1 - E_3^2 & E_2 E_3 - E_1 \\
E_3 E_1 - E_2 & E_1 (1 - E_3^2) & 1 - E_1^2 
\end{array}\right) \,.
\ee
 The equations display a cyclic symmetry in the indices $1,2,3$ which is also apparent in 
the matrix $\bA$. The cyclic symmetry results from the nature of the problem since we are 
free to choose the labeling of the three space-crafts. In the matrix $\bA$ we must also 
change the rows/columns consistently performing the cyclic change of the indices. The 
cyclic symmetry is further carried over to the solutions $(\bp, \bq)$.
\par 
The integration of the eq. (\ref{simpeq}) can be  
carried out in time-steps of $s$. The integration is immediate if we
operate the eq. 
(\ref{simpeq}) on $\bV$. We first need to take the transpose of eq. (\ref{simpeq}) and 
then operate on $\bV$. We then obtain:
\bea
\D \bp^T \bV &=& \bq^T \bA^T  \ \bV \no \\
           &=& \bq^T (\bmu_{adj} \bmu^T)^T \  \bV \no \\
           &=& \bq^T \bmu \D \  \bC \no \\
           &=& - \D \bq^T \  \bU \, ,
\eea
which gives,
\be
\D (\bp^T \ \bV + \bq^T \  \bU) = 0.
\ee  
This equation immediately integrates to,
\be
(\bp^T \ \bV + \bq^T \ \bU) (t + ns) = (\bp^T \ \bV + \bq^T
\ \bU) (t),
\ee    
where $n$ is an integer. If we arbitrarily set $t=0$ and if 
$(\bp^T \ \bV + \bq^T \ \bU) (0) = 0$, then $(\bp^T \ \bV +
\bq^T \ \bU) (ns) = 0$.  
\par
It is not clear to us, how the above solution would be useful physically, but we present 
it as an interesting outcome. However, the main problem is of seeking solutions 
$(\bp, \bq)$ to eq.(\ref{opeq}). This problem and its solution we discuss in the next 
section.

\section{The modules of syzygies}

Several solutions have been exhibited in \cite{AET99,ETA00} to 
eq.(\ref{opeq}). The solutions 
have the characteristic property that the $\det \bmu$ cancels on both sides leading to polynomial 
vectors $\bp$ and $\bq$. We reproduce here the solutions obtained in previous works. 
The solution $\zeta$ is given by $\bp^T = \bq^T = (E_1, E_2, E_3)$. The solution 
$\alpha$ is described by $\bp^T = (1, E_3, E_1 E_3)$ and $\bq^T = (1, E_1 E_2, E_2)$. 
The solutions $\beta$ and $\gamma$ are obtained from $\alpha$ by cyclically permuting 
the indices of $E_k, \bp$ and $\bq$. These solutions as realised in earlier works 
are important, because they consist of polynomials with lowest possible degrees and 
thus are simple. Other solutions containing higher degree polynomials can be 
generated conveniently from these solutions. Linear combinations of these solutions 
are also solutions to the given system of equations. But it 
is not clear that this procedure generates all the solutions. In particular, $\zeta$ 
cannot be generated from the set $\alpha,\ \beta$ and $\gamma$,  
where generating a data combination means writing it as a linear combination of 
the elements belonging to the generating set.
\par
The basic reason, as mentioned earlier, is that we do not have a vector space here. 
Three independent constraints on a six tuple do not produce a space which is 
necessarily generated by three basis elements. This 
conclusion would follow if the solutions formed a vector space but they do not. 
The polynomial six-tuple $\bp, \bq$ can be multiplied by polynomials in 
$E_1, E_2, E_3$ (scalars) which do not form a field. So that the 
inverse in general does not exist within the ring of polynomials. We therefore have a 
module over the ring of polynomials in the three variables $E_1, E_2, E_3$. 
\par
In this section we present the general methodology for obtaining the solutions to 
(\ref{opeq}). The method is illustrated by applying it to equations (\ref{opeq}). 
In the next subsection we consider the more general problem of optical bench motions as 
well. The optical bench motion noise can also be eliminated using the same method. 

\subsection{Cancellation of laser frequency noise}

There are three linear constraints on the polynomials given by the equations (\ref{opeq}). 
Since the equations are linear the solutions space is a submodule of the module of 
six-tuples of polynomials. The module of six-tuples is a free module, i.e. it has six 
basis elements that not only generate the module but are linearly independent. A
natural choice of the basis is $f_i = (0, ..., 1, ...,0)$ with 1 in the $i$-th place 
and 0 everywhere else; $i$ runs from 1 to 6. The 
definitions of generation (spanning) and linear independence are the same as that for vector 
spaces. A free module is essentially like a vector space. But our interest lies in its 
submodule which need not be free and need not have just three generators as it would 
seem if we were dealing with vector spaces.
\par
The problem at hand is of finding the generators of this submodule i.e. any element of 
the module should be expressible as a linear combination of the generating set. 
In this way the generators are capable of spanning the full module or generating the 
module. We examine the eq.({\ref{opeq}) explicitly componentwise:
\bea
p_1 - q_1 + E_3 q_2 - E_2 p_3 &=& 0, \no \\
p_2 - q_2 + E_1 q_3 - E_3 p_1 &=& 0, \no \\
p_3 - q_3 + E_2 q_1 - E_1 p_2 &=& 0.
\label{lneq}
\eea
The first step is to use Gaussian elimination to obtain $p_1$ and $p_2$ in terms of 
$p_3, q_1, q_2, q_3$. We then obtain:
\bea
p_1 &=& q_1 - E_3 q_2 + E_2 p_3, \no \\
p_2 &=& q_2 - E_1 q_3 + E_3 p_1 \no \\
    &=& E_3 q_1 + (1 - E_3^2) q_2 - E_1 q_3 + E_2 E_3 p_3, 
\label{gauss}
\eea
and then substitute these values in the third equation to obtain a linear implicit 
relation between $p_3, q_1, q_2, q_3$. We then have:
\be
(E_1 E_2 E_3 - 1) p_3 + (E_1 E_3 - E_2) q_1 + E_1 (1 - E_3^2) q_2 + (1 - E_1^2) q_3 = 0. 
\label{implct}
\ee
Obtaining solutions to (\ref{implct}) amounts to solving the problem since the  
the remaining polynomials $p_1, p_2$ have been expressed in terms of  $p_3, q_1, q_2, q_3$
in (\ref{gauss}).
\par
The solutions to (\ref{implct}) form the {\it first module of syzygies} of the 
coefficients in the eq. (\ref{implct}), namely,
 \\
$E_1 E_2 E_3 - 1,E_1 E_3 - E_2, E_1 (1 - E_3^2), 1 - E_1^2$. 
The generators of this 
module can be obtained from standard methods available in the literature. We briefly 
outline the method given in the books by Becker et al. \cite{becker} and 
Kreuzer and Robbiano \cite{martin} below. The details have been included in the Appendix.  

\subsubsection{Groebner basis}

The first step is to obtain the Groebner basis for the ideal $\cU$ generated by the 
coefficients:
\be 
u_1 = E_1 E_2 E_3 - 1,~~u_2 = E_1 E_3 - E_2,~~u_3 = E_1 (1 - E_3^2),~~u_4 = 1 - E_1^2. 
\label{idgen}
\ee
The ideal $\cU$ consists of linear combinations of the form $\sum v_i u_i$ where $v_i$, 
$i = 1, ..., 4$ are polynomials in the ring ${\cal Q}[E_1, E_2, E_3]$ where ${\cal Q}$ 
is the field of rational numbers. There can be several sets of generators
for $\cU$. 
A Groebner basis is a set of generators which is `small' in a specific sense. 
\par
There are several ways to look at the theory of Groebner basis. One way is, suppose we are 
given polynomials $g_1, g_2, ..., g_m$ in one variable over say ${\cal Q}$ and we would 
like to know whether another polynomial $f$ belongs to the ideal generated by the 
$g$'s. A good way to decide the issue would be to first compute the gcd (greatest common 
divisor) $g$ of $g_1, g_2, ..., g_m$ and checking whether $f$ is a multiple of $g$. One can 
achieve this by doing the long division of $f$ by $g$ and checking whether the remainder 
is zero. All this is possible because ${\cal Q}[x]$ is a Euclidean domain and also a 
principle ideal domain (PID) wherein any ideal is generated by a single element. Therefore 
we have essentially just one polynomial - the gcd - which generates the ideal generated by 
$g_1, g_2, ..., g_m$. The ring of integers or the ring of polynomials in one variable over 
any field are examples of PIDs whose ideals are generated by single elements.
However, when we consider more general rings (not PIDs) like the one we are 
dealing with here, we do not have a single gcd but a set of several polynomials which 
generates an ideal in general. A Groebner basis of an ideal can be thought of as a 
generalisation of the gcd. In the univariate case, the Groebner basis reduces to the gcd.
\par
Groebner basis theory generalises these ideas to multivariate polynomials which are neither 
Euclidean rings nor PIDs. Since there is in general not a single generator for an ideal, 
Groebner basis theory comes up with the idea of dividing a polynomial with a {\it set} 
of polynomials, the set of generators of the ideal, so that by successive divisions by 
the polynomials in this generating set of the given polynomial, the remainder becomes 
zero. Clearly, every generating set of polynomials need not possess this property. 
Those special generating sets that do possess this property (and they exist!) are 
called Groebner bases. In order for a division to be carried out in a sensible manner, 
an order must be put on the ring of polynomials, so that the final remainder after 
every division is strictly smaller than each of the divisors in the generating set. 
A natural order exists on the ring of integers or on the polynomial ring ${\cal Q}(x)$; 
the degree of the polynomial decides the order in ${\cal Q}(x)$. However, even for polynomials in two variables 
there is no natural order apriori ( Is $x^2 + y$ greater or smaller than $x + y^2$? ). 
But one can, by hand as it were, put a order on such a ring by saying $x >> y$, where 
$ >> $ is an order, called the lexicographical order. We follow this type of order,  
$E_1 >> E_2 >> E_3$  and ordering polynomials 
by considering their highest degree terms. It is possible to put different orderings on a 
given ring which then produces different Groebner bases. Clearly, a Groebner basis must 
have `small' elements so that division is possible and every element of the ideal 
when divided by the Groebner basis elements leaves zero remainder, i.e. every element 
modulo the Groebner basis reduces to zero.      
\par
In the literature, there exists a well-known algorithm called the the Buchberger 
algorithm which may be used to obtain the Groebner basis for a given set of polynomials 
in the ring. So a Groebner basis of $\cU$ can be obtained from the generators $u_i$ 
given in eq.(\ref{idgen}) using this algorithm. It is essentially again a generalisation 
of the usual long division that we perform on univariate polynomials. 
More conveniently, we prefer to use the wellknown `Mathematica' package. Mathematica 
yields a 3 element Groebner basis ${\cal G}$ for $\cU$:
\be
{\cal G} = \{E_3^2 - 1, E_2^2 - 1, E_1 - E_2 E_3 \} \, .
\ee
One can easily check that all the $u_i$ of eq.(\ref{idgen}) are linear combinations of 
the polynomials in ${\cal G}$ and hence ${\cal G}$ generates $\cU$. One also observes 
that the elements look `small' in the order mentioned above. However, one can 
satisfy oneself that ${\cal G}$ is a Groebner basis by using the standard methods 
available in the literature. One method consists of computing the S-polynomials (see 
Appendix A) for all the pairs of the Groebner basis elements and checking whether these 
reduce to zero modulo ${\cal G}$.
\par 
This Groebner basis of the ideal $\cU$ is then used to obtain the generators for the 
module of syzygies.

\subsubsection{Generating set for the module of syzygies}
 
The generating set for the module is obtained by further following the procedure
in the literature \cite{becker,martin}. The details are given in the Appendix 
specifically for our case.  We obtain 7 generators for the module. 
These generators do not form a minimal set and there are relations between them; in fact 
this method does not guarantee a minimum set of generators. These generators can be 
expressed as linear combinations of $\alpha, \beta, \gamma, \zeta$ and also in terms 
of $X^{(1)}, X^{(2)}, X^{(3)}, X^{(4)}$ given below in eq.(\ref{gen6}). The importance 
in obtaining the 7 generators is that the standard theorems guarantee that these 7 
generators do infact generate the required module.
Therefore from this proven set of generators we can check whether a particular set is 
infact a generating set. We present several generating sets below: 
\par
Alternatively, we may use a software package called `Macaulay 2' which calculates the 
generators given the the equations (\ref{lneq}). Using Macaulay 2, we obtain six 
generators. Again, Macaulay's algorithm does not yield a minimal set; we can express 
the last two generators in terms of the first four. Below we list this 
smaller  set of 
four generators in the order $X = (p_1, p_2, p_3, q_1, q_2, q_3)$:
\bea
X^{(1)} &=& (E_1 E_3 - E_2, 0, E_3^2-1, 0, E_2 E_3 - E_1, E_3^2 - 1), \no \\
X^{(2)} &=& (E_1, E_2, E_3, E_1, E_2, E_3), \no \\
X^{(3)} &=& (1, E_3, E_1 E_3, 1, E_1 E_2, E_2), \no \\
X^{(4)} &=& (E_1 E_2, 1, E_1, E_3, 1, E_2 E_3). 
\label{gen6}
\eea
Note that the last three generators are just $X^{(2)} = \zeta, X^{(3)} = \alpha, 
X^{(4)} = \beta$. But there is an extra generator $X^{(1)}$ needed to generate all 
the solutions.
\par
Another set of generators which could be useful for further work is a Groebner basis 
of a module. The concept of a Groebner basis of an ideal can be extended to that of 
a Groebner basis of a submodule of $(K[x_1, x_2, ..., x_n])^m$ where $K$ is a field, 
since a module over the polynomial ring can be considered as generalisation of an ideal 
in a polynomial ring. Just as in the case of an ideal, a Groebner basis for a module 
is a generating set with special properties. For the module under consideration we 
obtain a Groebner basis using Macaulay 2 : 
\bea
G^{(1)} &=& (E_1, E_2, E_3, E_1, E_2, E_3), \no \\
G^{(2)} &=& (E_1 E_3 - E_2, 0, E_3^2-1, 0, E_2 E_3 - E_1, E_3^2 - 1), \no \\
G^{(3)} &=& (E_1 E_2, 1, E_1, E_3, 1, E_2 E_3), \no \\
G^{(4)} &=& (1, E_3, E_1 E_3, 1, E_1 E_2, E_2), \no \\
G^{(5)} &=& (E_3 (E_1^2 - 1), 1 - E_3^2, 0, 0, 1 - E_1^2, E_1 (E_3^2 - 1)). 
\label{grbn6}
\eea  
Note that in this Groebner basis $G^{(1)} = \zeta = X^{(2)}, G^{(2)} = X^{(1)}, 
G^{(3)} = \beta = X^{(4)}, G^{(4)} = \alpha = X^{(3)}$. Only $G^{(5)}$ is the new generator. 
\par
Another set of generators are just $\alpha, \beta, \gamma$ and $\zeta$. This can be 
checked using Macaulay 2 or one can relate $\alpha, \beta, \gamma$ and $\zeta$ to the 
generators $X^{(A)}, A = 1, 2, 3, 4$ by polynomial matrices. In the Appendix, we express 
the 7 generators we obtained following the literature, in terms of 
$\alpha, \beta, \gamma$ and $\zeta$. Also we express $\alpha, \beta, \gamma$ and $\zeta$ 
in terms of $X^{(A)}$. This proves that all these sets generate the 
required module of syzygies.
\par
The question now arises as to which set of generators we should choose 
which facilitates  further analysis. The analysis is simplified 
if we choose a smaller number of generators. Also we would prefer low degree
polynomials to appear in the generators so as to avoid  cancellation of 
leading terms in the polynomials. By these two criteria we may choose, 
$X^{(A)}$ or $\alpha, \beta, \gamma, \zeta$. Among these two sets of
generators, we arbitrarily make the choice of $X^{(A)}$.
     
\subsection{Cancellation of noise from moving optical benches}

The work done in \cite{AET99,ETA00} can be conveniently reexpressed in our formulation and 
leads to further insights into the problem. 
\par
There are two optical benches on each space-craft which have random velocities and are 
connected by optical fibers. The random velocities of the optical benches and the 
delay in the optical fibers are measured as further Doppler shifts apart from 
other noise and the GW signal. Since we are interested in the cancellation of 
laser frequency noise and motion of the optical benches, we write expressions for 
the beams containing only these quantities. The Doppler beams will of course contain 
other effects arising from shot noise, GW signal, motion of proof masses etc., but we 
will not write them in the expressions for the Doppler data because they are not 
relevant to the problem we are interested in. We follow the notations of 
\cite{AET99,ETA00}. The quantities 
pertaining to the left bench will be unstarred while that for the right bench are 
starred. There are now twelve Doppler data streams which need to be combined in 
an appropriate manner in order to cancel the noise from the laser as well as from 
the motion of the optical benches. The fractional frequency fluctuations of laser on 
the left optical bench $i$ are denoted by $C_i$ and on the right optical bench $i^*$ 
by $C_i^*$, the random velocities of the 
benches $\bV_i, \bV_i^*$ and $\eta_i$ the frequency shifts in the optical fibers 
connecting the optical benches in space-craft $i$. We then have the following expressions 
for the four data streams pertaining to space-craft $1$:
\bea
U^1 &=& E_2 (C_3 - \un_2 \cdot  \bV_3 ) - (\un_2  \cdot  \bV^*_1 + C_1^* ), \no \\
V^1 &=& - E_3 (C_2^* + \un_3 \cdot \bV_2^* ) + ( C_1 - \un_3 \cdot \bV_1 ), \no \\
z_1 &=& C_1 - C_1^* + \eta_1 - 2 \un_3 \cdot \bV_1\, ,  \no \\
z_1^* &=& C_1^* - C_1 + \eta_1 + 2 \un_2 \cdot \bV_1^* \, .  
\eea  
The other eight data streams on space crafts 2 and 3 are obtained by cyclic permutations 
of the indices in the above equations. Here $\un_2$ denotes a unit vector in the direction 
from space-craft 1 to space-craft 3 and the remaining unit vectors $\un_3$
and $\un_1$ are 
obtained by cyclically permutating the indices. 
\par
We find that the twelve Doppler data streams depend only on the  particular combinations 
$C_1 - \un_3 \cdot  \bV_1$ and $C_1^* + \un_2 \cdot  \bV_1^*$ and their cyclic permutations. We 
define these combinations as $\tC_1$ and $ \tC_1^*$ respectively, i.e.:
\bea
\tC_1 &=& C_1 -  \un_3 \cdot  \bV_1 \, , \no \\
\tC_1^* &=&  C_1^* + \un_2 \cdot  \bV_1^* \, , 
\eea
and also their cyclic permutations. Then the expressions for the data streams simplify 
considerably. We write the expressions for the data streams corresponding to space-craft 1.
Others are obtained as before by cyclic permutations. 
\bea
U^1 &=& E_2 \tC_3 - \tC_1^* \, , \no \\
V^1 &=& -E_3 \tC_2^* + \tC_1\, ,  \no \\
Z^1 &=& {1 \over 2} (z_1 - z_1^*) \no \\
    &=& \tC_1 - \tC_1^*\, . 
\eea
The new variables $Z^i$ have been defined which automatically cancel the $\eta_i$. Also 
we note that the $U^i, V^i$ have the same form as in eq. (\ref{beams}), except that the 
$C_i$ are replaced by the $\tC_i$ which account for the motions of the optical benches. 
\par 
The noise cancellation condition now becomes:
\be
p_i V^i + q_i U^i + r_i Z^i = 0 \, , 
\ee 
where $r_i$ are polynomials in $E_1, E_2, E_3$. Since the above equations should hold for 
any $\tC_i, \tC_i^*$, we obtain the following equations the polynomials $p_i, q_i, r_i$ must 
satisfy:
\bea
p_1 + E_3 q_2 + r_1 &=& 0 \, , \no \\
E_2 p_3 + q_1 + r_1 &=& 0 \, , \no \\
p_2 + E_1 q_3 + r_2 &=& 0 \, , \no \\
E_3 p_1 + q_2 + r_2 &=& 0 \, , \no \\
p_3 + E_2 q_1 + r_3 &=& 0 \, , \no \\
E_1 p_2 + q_3 + r_3 &=& 0 \,  .
\label{obeq}
\eea
We now have a nine component polynomial vector. The solutions to (\ref{obeq}) form another 
module of syzygies which is related in a simple way to the module obtained in 
just laser noise cancellation. Eliminating $r_i$ from eq. (\ref{obeq}) we obtain the 
same equations for $p_i$ and $q_i$ as in (\ref{lneq}). This enables us to extend the 
previously obtained generating sets to this module. Thus, thanks to the mapping 
of $C_i, ( C_i^* ) \longrightarrow \tC_i, ( \tC_i^* )$,  the two modules
are isomorphic. We just state the remaining three 
entries ($r_1, r_2, r_3$) of the generating sets keeping the same notation. The first 
set of 4 generators in the order $(r_1, r_2, r_3)$ are:
\bea
X^{(1)} &=& (E_2 (1 - E_3^2), E_1 (1 - E_3^2), 1 - E_3^2)  \, ,\no \\
X^{(2)} &=& - (E_1 + E_2 E_3, E_2 + E_1 E_3, E_3 + E_1 E_2) \, , \no \\
X^{(3)} &=& - (1 + E_1 E_2 E_3, E_1 E_2 + E_3, E_1 E_3 + E_2) \, , \no \\
X^{(4)} &=& - (E_1 E_2 + E_3, 1 + E_1 E_2 E_3, E_1 + E_2 E_3) \, . 
\label{res}
\eea
In the other generating set, namely, the Groebner basis we need to specify just  
$G_5$ since the other elements are in the previous generating set. Thus,
\be
G^{(5)} = (0, (E_1^2 - 1)(1 - E_3^2), 0).
\ee   
%
\section{The Detector  response}
%
The ripples produced in the spacetime  by the gravitational waves  
as they propagate through the LISA detector are measured as the Doppler shifts in the
laser frequency.  The measured signals will have the various 
noises along with the Doppler shift produced by the gravitational radiation.
In the last section we have studied various combinations 
of beams which cancels the laser phase noise and optical bench 
acceleration noise. In this  section  we investigate the  response
of the detector  for these combinations. We compute the transfer 
functions for the generators and also their linear combinations.
The laser phase noise and optical bench acceleration
noise is then also cancelled for the linear combinations. However, noises such 
as the shot noise and the acceleration noise of the proof masses do  not cancel out. In 
the following subsections we set up the coordinate system adapted to the LISA 
geometry and then go on to compute the response of LISA. 
\subsection{ Parameterisation of the interferometer}
Figure \ref{fig1} describes the LISA configuration. We choose a coordinate system  
in which the LISA triangle is at rest. Although this 
coordinate system is in motion relative to the usual coordinate systems 
normally encountered, we will find such a system of coordinates convenient for 
further analysis, such as computing SNR's of monochromatic sources etc.
  
	The unit vector  $\hat{w}$ connecting the origin and the
	source is parameterised
	by the source angular  location $(\theta, \ \phi )$, so that
\begin{equation}
\hat{w} \ = \ \left( \begin{array}{c}
		\sin\theta \ \cos\phi \\
		\sin\theta \ \sin\phi \\
		\cos\theta
		\end{array}
		\right) \, ,
\end{equation}
the transverse plane is spanned by the unit transverse vector $\hat{\theta}$
and $\hat{\phi}$, defined by
\begin{eqnarray}
\hat{\theta} \ = \ \frac{ \partial \hat{w}}{  \partial \theta},
\ \ \ \  \hat{\phi} \ = \  \frac{1}{\sin\theta}  \frac{ \partial \hat{w}}{  \partial 
\phi} \, .
\end{eqnarray}
As the wave propagates through the LISA triangle, the 
components of the gravitational perturbation can be written as
\begin{equation}
h_{ij}(t, \vec{r})\ = \ h_{+} \left( t- \hat{w} \cdot \vec{r}  \right)
\left( \theta_i \theta_j - \phi_i \phi_j  \right)
\ + \
 h_{\times}  \left( t- \hat{w} \cdot \vec{r}  \right)  
\left( \theta_i \phi_j  + \theta_j \phi_i 
\right) \, , \label{HIJ}
\end{equation}
where $h_+$ and $h_{\times}$ are arbitrary  functions describing the two 
GW amplitudes. 
 
We consider the effect of this perturbation on the
light beam traveling between two points A and B. From this we
 obtain the complete response.
Let $\vec{r}_A$ and $\vec{r}_B$ be the position vectors of points A and B respectively.   
Then the line element of the spacetime along the beam {null ray} obeys,
\begin{equation}
0 \ = \ dt^2  -  dx^2  -  dy^2  -  dz^2  +  h_{ij} dx^i dx^j ,
\label{NULL}
\end{equation}
where the $i,j$ run over space indices only.
If $ \hat{n}$ is the unit vector directed from A to B, we have
\begin{equation} 
dx^i \ = \ n^i d\lambda \, ,
\end{equation}
where $\lambda$ is the Euclidean length. The equations (\ref{NULL})
can be expressed as,
\begin{equation}
0 \ = \ dt^2 - d\lambda^2 \left[ 1 - h_{ij} n^i n^j \right] \, ,
\end{equation}
or equivalently 
\begin{equation} 
d\lambda \ = \ dt \left[ 1 + \frac{1}{2}h\left( t - \hat{w} \cdot
\vec{r} \right) \right] \, .
\end{equation} 
From eq. (\ref{HIJ}) we get 
\begin{equation}
h(t) \ = \ h_+(t) \xi_+(\theta, \phi) + h_{\times}(t) \xi_{\times}
(\theta, \phi) \, ,
\end{equation} 
\begin{equation}
\left\{  \begin{array}{l}
\xi_+ \ = \ (\hat{\theta}\cdot \hat{n})^2 - (\hat{\phi}\cdot
\hat{n})^2 \\
\xi_{\times} \ = \ 2 (\hat{\theta}\cdot\hat{n})(\hat{\phi}\cdot\hat{n})
\end{array}
\right. \, .
\end{equation} 

%
\subsection{$h$-sensitivity of one arm}
We now apply the above analysis to compute the Doppler response of the laser beam 
along one arm of the LISA detector.
Let the beam start at $t=t_0$ from the point $\vec{r}_A$   
and travel towards the  point $\vec{r}_B$ and reach it at  $t=t_1$. Then,
\begin{equation}
\vec{r}(t)  \ = \ \vec{r}_A + (t -t_0 ) \hat{n}, \ \ \  
\vec{r}(t_1) \ = \  \vec{r}_B \, .
\end{equation}
The line element along this path satisfies the equation,
\begin{equation}
d\lambda \ = \ dt\left\{ 1 + \frac12 h\left[ ( 1- \hat{w} \cdot \hat{n})t 
- \hat{w} \cdot \vec{r}_A + t_0  \hat{w} \cdot \hat{n} \right]
  \right\} \, .
\end{equation} 
The global travel time $t_1 - t_0 $ is given by the integral :
\begin{equation}
L\ = \ t_1 - t_0 + \frac12 \int_{t_0}^{t_1} 
 h\left[t ( 1- \hat{w} \cdot \hat{n})
 - \hat{w} \cdot \vec{r}_A + t_0  \hat{w} \cdot \hat{n} \right]
dt \, .
\end{equation} 
It is convenient for many purposes to pursue our analysis in the Fourier domain. 
We Fourier transform the GW amplitude $h$:
\begin{equation}
h(t) \ = \ \int d \Omega \ \tilde{h}(\Omega) exp(- i \Omega t ) \, ,
\end{equation} 
 and  the travel time can be expressed as,
\begin{equation}
t_1-t_0 \ = \ L - \frac12 \int d\Omega \tilde{h}(\Omega)\ \int_{t_0}^{t_1}
exp\left[ -i \Omega \left( 1 - \hat{w} \cdot \hat{n} \right)t \right]
exp\left[ i \Omega \hat{w} \cdot \vec{r}_A \right]
exp\left[ -i  \Omega t_0 \hat{w} \cdot \hat{n}\right] dt \, .
\end{equation}  

In the zeroth order of the integral, we have $t_1 - t_0 \ = L $, and we obtain:
\begin{eqnarray} 
t_1- t_0 & = & L   - \frac12 \int d\Omega \tilde{h}(\Omega) \ exp\left( i
\Omega \hat{w} \cdot
\vec{r}_A \right) 
exp\left( -i  \Omega t_1\right)  \\
\ & \times & \frac{ exp\left( i \Omega L \hat{w} \cdot  \hat{n}\right)
- exp( i \Omega L) }{ -i \Omega ( 1 - \hat{w} \cdot \hat{n})}    \, . 
\end{eqnarray} 
The phase change over that time interval is $\Phi=\omega(t_1 -t_0)$,
where $\omega= 2 \pi \nu_{opt}$ is the optical circular frequency. We
can assume that the time $t_1$ is the current time and $t_0$ the retarded time,
so that  the phase is $\Phi(t)=\omega(t-t_0)$: 
\begin{eqnarray}
\Phi(t)\  = \  \omega L &  - & \frac{\omega}{2( 1- \hat{w} \cdot \hat{n})}
\int  d\Omega \tilde{h}(\Omega) \ exp\left( i
\Omega \hat{w} \cdot
\vec{r}_A \right)
exp\left( -i  \Omega t\right)  \\
\ & \times & \frac{ exp\left( i \Omega L \hat{w} \cdot  \hat{n}\right)
- exp( i \Omega L) }{ -i \Omega }  \, .
\end{eqnarray} 
By taking the time derivative, we get the instantaneous frequency, 
\begin{equation}
\frac{ \delta \nu(t)}{\nu_{opt}} \ = \ \frac{1}{\omega} \ \frac{ d\Phi(t)}
{dt} \, .
\end{equation}
In the time domain and using $\vec{r}_A  + L \hat{n} \ = \ \vec{r}_B $
we finally get,
\begin{equation}
\frac{ \delta \nu(t)}{\nu_{opt}} \ = \  \frac{-1}{2( 1- \hat{w} \cdot \hat{n})}
 \left[
  h\left( t - \hat{w} \cdot
   \vec{r}_B\right)  - h \left( t -\hat{w}\cdot  \vec{r}_A - L \right) \right].
\end{equation} 
In the Fourier domain we may express this result as:
\begin{equation}
\frac{ \widetilde{\delta \nu(\Omega)}}{\nu_{opt}} \ = \ 
\frac{\tilde{h}(\Omega)}{2( 1- \hat{w} \cdot  \hat{n})}
exp\left[ i \Omega (L+ \hat{w} \cdot
 \vec{r}_A ) \right] \left( 1 - exp\left[ -i \Omega L ( 1 - \hat{w} \cdot
 \hat{n}  ) \right] \right) \, .
 \end{equation} 
\subsection{ $h$-sensitivity of the elementary Doppler data}
In this section we compute the expression for the transfer function
for the six elementary beams given in the eq. (\ref{beams}). These  
beams further can be combined with suitable delays as described in the previous 
sections for achieving cancellation of laser phase noise.

The Doppler shift being expressed in the Fourier domain as,
\begin{equation} 
\frac{ \widetilde{\delta \nu(\Omega)}}{\nu_{opt}} \ = \ 
\tilde{h}_+(\Omega) F_+(\Omega) + \tilde{h}_{\times}(\Omega) F_{\times}(\Omega)
\, ,
\end{equation} 
where $F_{+, \times}(\Omega)$ are transfer functions. We can compute the
transfer functions for  the combinations $U_i, \ V_i$. We just give below 
$F_{U_1;+,\times}$ and $F_{V_1;+,\times}$; the others are obtained by cyclic 
permutations:
\begin{eqnarray}
F_{U_1;+,\times} & = & \frac{e^{ i \Omega\left(\hat{w}\cdot\vec{r}_3 +
L_2\right)}} { 2 \left( 1 + \hat{w} \cdot \hat{n}_2 \right)} \ 
\left( 1 - e^{ - i \Omega L_2 \left( 1 + \hat{w} \cdot \hat{n}_2
\right)} \right) \xi_{2;+,\times} \, , \nonumber  \\
F_{V_1;+,\times} & = & -\frac{e^{ i \Omega\left(\hat{w}\cdot\vec{r}_2 +
L_3\right)}} { 2 \left( 1 - \hat{w} \cdot \hat{n}_3 \right)} \
\left( 1 - e^{ - i \Omega L_3 \left( 1 - \hat{w} \cdot \hat{n}_3
\right)} \right) \xi_{3;+,\times}  \, ,
\end{eqnarray} 
where,
\begin{equation}
\xi_{i;+} \ = \ (\hat{\theta}\cdot \hat{n}_i)^2 - (\hat{\phi}\cdot
\hat{n}_i)^2 \  \ \ \ \
\xi_{i;\times} \ = \ 2 (\hat{\theta}\cdot\hat{n}_i)(\hat{\phi}\cdot\hat{n}_i)
\, .
\end{equation}
To implement the cancellation of laser phase noise these elementary beams must 
be combined with suitable time-delays. We notice that in the Fourier domain 
the delay operators get replaced by simple multiplicative factors as the following 
computations show. This is one of the advantages of the Fourier analysis.  
The delay operators introduced in the section are $E_i$ such that for any
function of time $f(t)$, we have
\begin{equation}
E_i*f(t) \ = \ f(t-L_i) \, ,
\end{equation} 
which in the Fourier domain in nothing but
\begin{equation}
\widetilde{ E_i* f(\Omega)} \ = \  e_i(\Omega) \tilde{f}(\Omega) \, ,
\end{equation} 
where the $e_i$ are simple factors:
\begin{equation}
 e_i \ = \ e^{i \Omega L_i} \, .
\end{equation} 
Thus operator polynomials in $E_i$ become actual polynomials in $e_i$ in the 
Fourier domain. This particularly simple fact can be used to advantage for the 
simple but astrophysically important sources, namely the monochromatic sources 
considered in the next section.
\par
Below in the figures 2(a) and 2(b) we present the transfer functions $F _{V_{1}}$ for 
both polarisations: 
\begin{figure}[t]
\epsfig{file=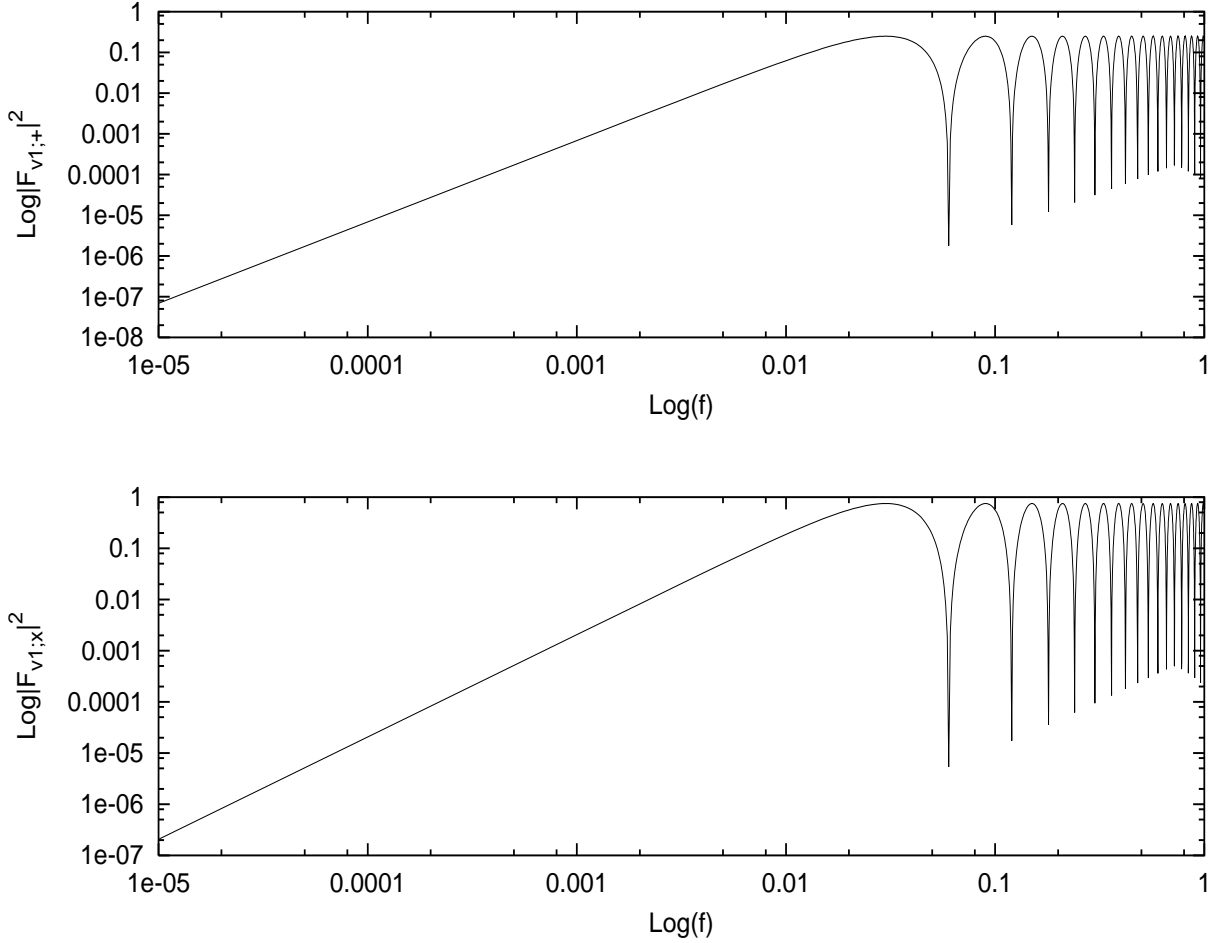,height=13cm,width=17cm}
\caption{ Log Log plot of the psd 
$|F_{V_{1;+}}|^2 $  and  $|F_{V_{1;\times}}|^2 $,  of the corresponding 
transfer function  are displayed in (a) and (b) repsectively,  as a 
function of frequency 
for $\theta=0$, $\phi=0$. }
\label{fig3}
\end{figure}
The other transfer functions show similar characterstics.

\section{Monochromatic sources}
\subsection{Noise}
We recall that the laser frequency noise and optical bench motion 
noise can be cancelled by taking appropriate combinations of the beams. In this 
scheme the noises that do not  cancel out in the module of {\em syzygies}
are the acceleration noise of the proof masses and the shot noise. These then 
form the bulk of noise spectrum which we obtain below for any given 
data combination $X$. We compute the noise power spectral densities for the
generators $X^{(A)}$.

The beam with the signal and the various noises can
be written as,
\begin{eqnarray}
U^1 & = & E_2 \tilde{C}_3 - \tilde{C}^*_1 + 2 \hat{n}_2 \cdot \vec{v}^*_1
+ h_{U1} + Y^{shot}_{U1} \\
V^1 & = & -E_3 \tilde{C}^*_2 + \tilde{C}_1 + 2 \hat{n}_3 \cdot \vec{v}_1
- h_{V1} - Y^{shot}_{V1} \\
Z^1 & = & \left ( \frac{1}{2} \right )( z_1 - z_1^*) + \hat{n}_3 \cdot \vec{v}_1 +
\hat{n}_2 \cdot \vec{v}^*_1 \, .
\end{eqnarray}
The other beams can be obtained by taking cyclic permutations. Here
$\vec{v}_1$ and $\vec{v}^*_1$  are the random velocities of the proof
masses, in the left and right branches respectively, in space craft 1.

Let the noise cancelling combination $X$ be given by,
\begin{equation}
X=p_iV^i + q_i U^i + r_i Z^i \, , \label{COM1}
\end{equation}
where the 9-tuple $(p_i, q_i, r_i)$ is in the module of syzygies.
This combination cancels laser phase noise and the optical bench 
acceleration noise, where as the shot noise and proof masses acceleration
noise do not cancel. Using eq. (\ref{COM1})
we obtain the power spectral density of X for the two noises,
\begin{equation}
<X^2>^{proofmass} \ \ = \sum_{i=1}^3 \left( |2 p_i + r_i|^2 + 
| 2 q_i + r_i |^2 \right) S^{proofmass} \, ,
\label{accnoise}
\end{equation} 
\begin{equation}
<X^2>^{shotnoise} \ \  = \sum_{i=1}^3 \left( |p_i|^2 +
|q_i|^2 \right) S^{shotnoise}\, ,
\label{shnoise}
\end{equation}

where $S^{proofmass}$ is obtained from $\vec{v}_i$ and $\vec{v}^*_j$. 
Here following the literature \cite{RIP} we take, 
$ S^{proofmass}= 2.5 \times 10^{-48} \left[ f/1 \ Hz\right]^{-2} Hz^{-1}$
and 
 $ S^{shotnoise}= 5.3 \times 10^{-38} \left[ f/1 \ Hz\right]^{2} Hz^{-1}$.

Explicit simplified expressions for the noise may be obtained by assuming,
$$
e_1 \ = \ e_2 \ = \ e_3  \ \equiv \  e^{i \Omega L}
$$
In the particular cases of the generators $X^{(A)}, A = 1, 2, 3, 4$ we obtain, 
\begin{eqnarray}
S_{X^{(1)}}(f) & = & \left[16 \sin^2( 2 \pi f L ) + 32 \sin^4 \pi f L ) \right] S^{proofmass} + \left[ 8 \sin^2(\pi f L ) + 8 \sin^2(2 \pi f L ) \right] 
 S^{shotnoise} \\
S_{X^{(2)}}(f) & = & 24 \sin^2(\pi f L) S^{proofmass} + 6  S^{shotnoise} \\
S_{X^{(3)}}(f)& = & \left[ 16 \sin^2(\pi f L) + 8 \sin^2(3 \pi f L ) \right]
 S^{proofmass} + 6 S^{shotnoise} \\
S_{X^{(4)}}(f) & = & \left[ 16 \sin^2(\pi f L) + 8 \sin^2(3 \pi f L ) \right]
 S^{proofmass} + 6 S^{shotnoise }
\end{eqnarray}
The plots of these noise spectra are shown in the figure~\ref{fig4}. 
\begin{figure}[t]
\epsfxsize=12cm
\begin{center}
\epsfbox{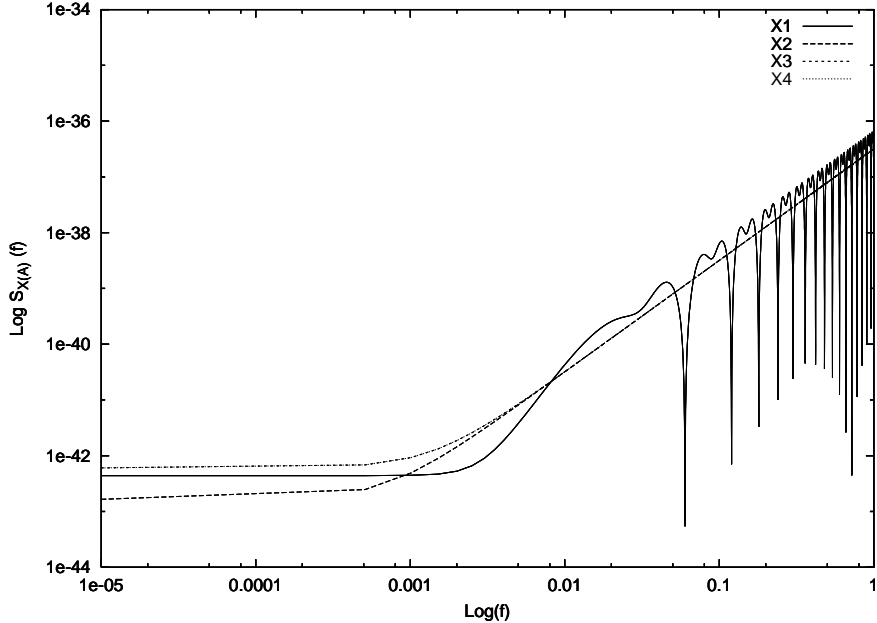}
\end{center}
\caption{Log Log plot of the noise spectra for the generators $X^{(A)}$. The curves 
for $X^{(3)}$ and $X^{(4)}$ coincide in the figure.
} \label{fig4}
\end{figure}
\subsection{Signal}
Monochromatic sources are  simplest among the gravitational
wave sources. There are a number of important objects such as¸
pulsars, rotating neutron stars, coalescing-binaries with 
sufficiently low mass may be considered as emitting monochromatic 
GW radiation. We will call those sources to be monochromatic which even if they 
change a little in frequency in a given observation time, the fractional change in 
SNR for the optimal data combination does not fall below a preassigned limit. We 
could take this limit to be few percent, but for concreteness we fix the limit at 1 $\%$.
The observation time $T$ we take to be one year. 
\par
For binary stars, the relevant quantity which decides the evolution in the GW 
frequency at a given frequency $f_0$ is the so called chirp mass 
${\cal M} = \mu^{3/5} M^{2/5}$, where $\mu$ is the reduced mass and $M$ is the total 
mass of the binary system. We assume a Newtonian evolution for the binary system which 
gives the rate of change of GW frequency $f$ as,
\be
{\dot f} = {3 \over 8} {f_0 \over \tau_c} ,
\ee 
where $\tau_c$ is the Newtonian coalescence time measured from the epoch when the 
GW frequency is $f_0$ and the `dot' denotes derivative with respect to time. 
The $\tau_c$ is given by,
\be
\tau_c \sim 2.15 \times 10^6 \left[ \frac{{\cal M}}{M_{\odot}}\right]^{-\frac{5}{3}} 
\left[\frac{ f_0}{1 {\rm mHz}} \right]^{-\frac{8}{3}}  {\rm years},
\ee
where $M_{\odot}$ is the solar mass.
\par
A  limit on the  rate of change of frequency can obtained  by considering 
the total change in the frequency $\Delta f$ during the period of observation $T$. 
That is,
\be
\Delta f = {\dot f} T .
\ee 
Inverting the above equations we obtain a limit on the chirp mass ${\cal M}$:
\begin{equation} 
{\cal M} \ \leq  175 \  \left[\frac{ f_0}
{1 {\rm mHz}} \right]^{-\frac{11}{5}} \ \left[ \frac{ \Delta f}{ 1 \mu Hz}
\right]^{\frac{3}{5}} M_{\odot} .
\end{equation} 
 In our investigation we take the bandwidth $ \Delta f$ 
by allowing $SNR$ to change  by 1\% at the frequency $f_0$. Numerically, we estimate 
$\Delta f$ for various values of $f_0$.   
The table below shows the upper bound for  ${\cal M}$  at various 
frequencies $f_0$:
\begin{center}
\begin{tabular}[t]{|c|c|c|}
\hline
$f_0$ in mHz &  $ \Delta f$ in $\mu$ Hz   &
$\hspace*{0.51cm}  \frac{{\cal M}}{M_{\odot}}$ \\
\ &  for 1\% change in SNR &   \\
\hline
0.1    &   1.0  & 27705 \\
1      &   9.9  & 691  \\
2      &   22    & 243   \\
10     &   1130   & 74  \\
\hline
\end{tabular}
\end{center}

Here our goal is to seek a data combination which 
optimises the SNR for a monochromatic source with given polarisation 
parameters and direction information. A convenient set of polarisation 
parameters  are the angles $\epsilon$ and $\psi$ describing the orientation 
of the angular momentum vector in space. The direction to the source is 
described by the  polar angles $\theta$ and $\phi$ in the coordinate 
system tied to LISA. 

For a monochromatic source the wave form can be written as
\begin{eqnarray}
h_+(t)&  =& \A \frac{ 1 + \cos^2 \epsilon}{2} \ \cos2\psi \ \cos( 2 \omega t) \, , \\
h_{\times}(t) & = & \A \cos\epsilon \  \sin2\psi \ \sin(2 \omega t) \, .
\end{eqnarray}
In the Fourier domain we have,
\begin{eqnarray}
h_+(\Omega)&  =& \A \ \frac{ 1 + \cos^2 \epsilon}{2} \ \cos2\psi \, , \\
h_{\times}(\Omega) & = & - i \A \ \cos\epsilon \  \sin2\psi \, .
\end{eqnarray} 
The response for the signal at the detector can now be written as,
\begin{eqnarray}
h_X & = & \sum_{i=1}^3 \left[ p_i \left( F_{Vi;+} h_+ + F_{Vi;\times}
h_{\times} \right) + q_i  \left( F_{Ui;+} h_+ + F_{Ui;\times} h_{\times}
\right) \right] \, ,
\end{eqnarray}
where $p_i$'s and $q_i$'s are  in the module of syzygies.
 From eq. (\ref{accnoise})
and (\ref{shnoise}) we can compute the total noise spectrum for 
the  generators  and  it can we written as,
\begin{eqnarray}
S_X(f)& = & \sum_{i=1}^3 \left[ \left( | 2 p_i + r_i |^2 + |2 q_i + r_i|^2
\right) S^{proofmass} + \left( |p_i|^2 + |q_i|^2\right) S^{shotnoise} \right] \, .
\end{eqnarray}
Expression for the signal to noise ratio (SNR)  for the  monochromatic source simplifies
to,
\be
SNR = \left\{ \frac{|h_X|}{\sqrt{S_X(f)}} \right\} \, .
\label{g_snr}
\ee
We plot the  sensitivities of the  generators $X^{(A)}$  as function of $f$  
in figures \ref{fig5}(a) and \ref{fig6}(a) by fixing the angles  
$\theta$ and $\phi$.  It is also important to understand the angular
dependence of the sensitivity of  generators, which are plotted in
the figures \ref{fig7}(a)-\ref{fig7}(d)  at a frequency of  $f=1$ mHz. The sensitivity $S$  
is defined following \cite{RIP},
\be
S = 5 {\sqrt{S_X  B} \over |h_X|} ,
\ee
where $B = 1/T$, where $T$ is the observation time which we take to be one year. The 
number 5 corresponds to an SNR of 5.
\ \\
\begin{figure}[t]
\begin{center}
\epsfbox{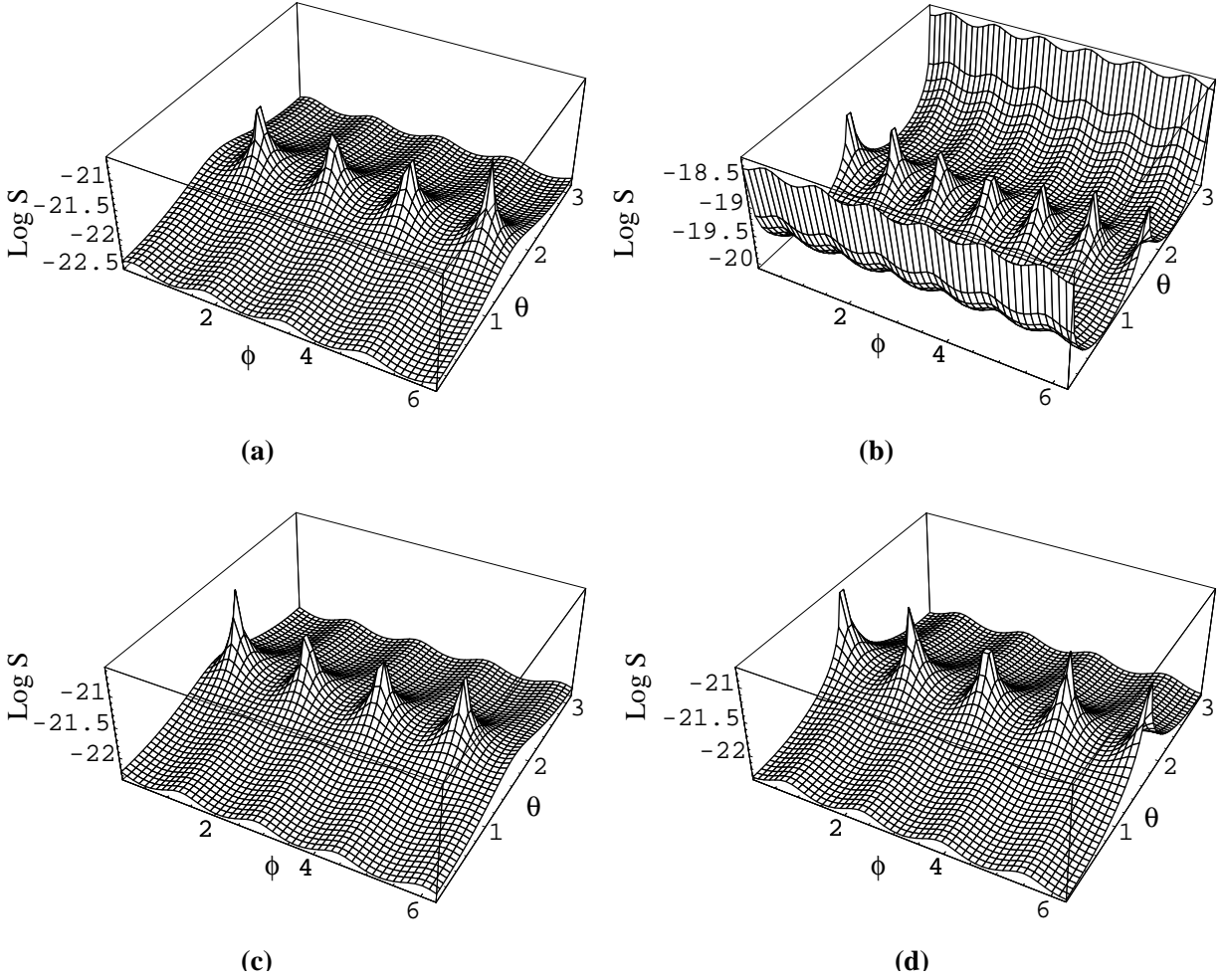}
\end{center}
\caption{Plots of $\log \ S$ of the generators 
$X^{(A)}, A = 1, 2, 3, 4$, are displayed in (a), (b), (c) and (d), respectively,
 as function of $\theta$ and  $\phi$ for $f= 1$ mHz and SNR = 5.
} \label{fig7}
\end{figure}   
\subsection{Maximisation }

 In this subsection our goal is to maximise the SNR for a given monochromatic source 
over the set of noise cancelling combinations. These combinations 
can be generated by the generators given in the eq. (\ref{gen6}) and (\ref{grbn6}). 
The SNR corresponding to the each of the generators ($X^{(A)}, \  A=1\ {\rm to \  }4$)  
as a function of frequency are shown in the figure. However one must maximise the 
SNR over an arbitrary linear combination of $X^{(A)}$. This goal is difficult to 
achieve since it involves a maximisation over a space of six tuples of polynomials 
which is essentially a function space. In order to make the problem tractable and still 
achieve adequate results we restrict the polynomials to be constants. This
approach  does 
not fully optimise the SNR but it comes quite close to  the 
 optimal solution. Our approach can be thought 
of as a zero'th order approximation. 
\par
A linear combination of the generators can be written as
\begin{equation}
	X \ = \ \sum_{A=1}^4  \alpha_{(A)}  X^{(A)},
\label{max_comb}
\end{equation}
here, $\alpha_{(A)} $ (for $A$= 1 to 4) are a set of real  numbers. Since a scalar 
multiple of $X$ will not yield anything new, we set one of the $\a$'s, say, 
$\a_{(1)} = 1$. Thus the SNR now becomes a function of three parameters $\a_{(i)}, i = 2, 3, 4$, 
which are just real numbers and our objective is to maximise the SNR with respect 
to $\a_{(i)}$. 
\par
In order to carry out the analysis efficiently and elegantly we find that it is useful 
to define  complex  noise vectors $N^{(A)}$ pertaining to $ X^{(A)}$ as follows:
\begin{equation}
N^{(A)} \ = \ \left( \sqrt{S_1} (2 p^{(A)}_i + r^{(A)}_i) ,  \sqrt{S_1}( 2 q^{(A)}_i + r^{(A)}_i ) , \sqrt{S_2}
p^{(A)}_i , \sqrt{S_2} q^{(A)}_i \right) 
\end{equation}
where, $p^{(A)}_i$,  $q^{(A)}_i$ and  $r^{(A)}_i$  corresponding to generators $X^{(A)}$
are given in the eq. (\ref{gen6}) and (\ref{res}) and \break
$S_1 \ = \  S^{proofmass}$ and  $S_2 \ = \  S^{shotnoise}$.
We have $N^{(A)}~~ \in  ~~C^{12}$ the 12 dimensional complex space and the usual scalar 
product $C^{12}$ induces a norm; $N^{(A)} \cdot {N^{(A)}}^* \equiv ||N^{(A)}||^2$ gives the noise 
psds corresponding to the basis $X^{(A)}$.

In a similar fashion one can also write the signal corresponding to a particular basis 
element. We first define the polynomial 6-tuple for each generator $X^{(A)}$ as follows:
\begin{equation}
P^{(A)} \ = \ \left(  p^{(A)}_i, q^{(A)}_i \right) \, ,
\end{equation}
and the GW signal 6-tuple as,
\begin{equation}
H^* = \left( F_{V_i;+} h_+ +  F_{V_i;\times} h_\times , \  F_{U_i;+} h_+ + 
  F_{U_i;\times} h_\times 
\right) .
\end{equation}
The signal for a specific generator $X^{(A)}$ is then written as,
\be
h^{(A)} =  P^{(A)}. H^*
\ee
and the corresponding SNR is given by,
\begin{equation}
SNR^{(A)} \  = \  \frac{|h^{(A)}|}{ ||N^{(A)}||}  \ = \ 
\sqrt{\frac{ (P^{(A)} \cdot H^* ) (P^{(A)} \cdot H^*)^*}
{N^{(A)} \cdot {N^{(A)}}^*}} \, .
\end{equation}
For an arbitrary linear combination X (eq. (\ref{max_comb})) the noise vector and
the signal vector can be expressed as
\be
N  =  \alpha_{(A)} N^{(A)}  ,~~~~~~
P  =  \alpha_{(A)} P^{(A)} \, , \label{comb}
\ee
where summation convention has been used. The signal is just the scalar product 
$h = P.H^* = \a_{(A)} h^{(A)}$. We omit subscripts $X$ on these quantities.

In this notation the SNR of the combination (\ref{max_comb}) can be written as,
\be
SNR \  = \ \frac{ |h|}{|| N ||}. 
\ee
Writing out explicitly the sums in the scalar products,
\be
(SNR)^2  \ = \ 
 \frac{ \alpha_{(A)} \alpha_{(B)} h^{(A)} {h^{(B)}}^* }
{\alpha_{(A)} \alpha_{(B)} N^{(A)} \cdot {N^{(B)}}^* }
\, .
\label{cals}
\ee
Maximisation with respect to $\a_{(2)}, \a_{(3)}, \a_{(4)}$ leads to the following three conditions 
which must be obeyed by $\a_{(i)}$ in order to yield the maximum SNR for $X$: 
\be
{\Re( h {h^{(i)}}^*) \over | h |^2 } = {\Re( N {N^{(i)}}^*) \over || N ||^2 },
\label{max_cond}
\ee
where $\Re(x)$ denotes the real part of the quantity $x$.

To demonstrate the usefulness of  the formalism, we consider just two generators 
$X^{(1)}$ and $X^{(2)}$. We take the $\alpha_{(1)} = 1 , \a_{(2)} = \a $ and other two 
$\a$ s zero. Then,
\be
X \ = \  X^{(1)} + \a  X^{(2)} \, .
\ee
The eq. (\ref{cals})  reduces to the form,
\be
{\cal S} \ = \ \frac{ a_1  + 2  b_1 \a +  c_1 \a^2 }{a_2  + 2 b_2 \a +  c_2
\a^2 } \, ,
\label{snr4_2}
\ee
where,
\bea
a_1 = |h^1|^2, ~~b_1 =\Re( h^1 h^{2*}),  ~~c_1=|h^2|^2 \, ,
\no \\
a_2 = |N^1|^2, ~~b_2 =\Re( N^1 N^{2*}),  ~~c_2=|N^2|^2 \, .
\eea
The condition for the optimisation (\ref{max_cond}) simplifies
to
\be
(b_1 a_2 - a_1 b_2 )+ (c_1 a_2 - a_1 c_2) \a + (c_1 b_2 - b_1 c_2 )  \a^2 \ = \ 0 \, .
\label{opt_al}
\ee
The two roots of the eq. (\ref{opt_al}) can be obtained.
Here, $\a$ is a function of the parameters $f$, $\theta$,
$\phi$, $\epsilon$ and $\psi$. 
One of the solutions of the eq. (\ref{opt_al})   correspond to
 the maximum and other correspond to the
minimum of the SNR. In a similar fashion one can  maximise the SNR by taking 
any two of the four generators given in the eq. (\ref{gen6}) and  by taking
appropriate $\alpha_{(i)}$ in the eq. (\ref{max_comb}). 
We have seen in several cases that maximising over just two generators
yields remarkably good results.

\begin{figure}[t]
\epsfxsize=12cm
\begin{center}
\epsfbox{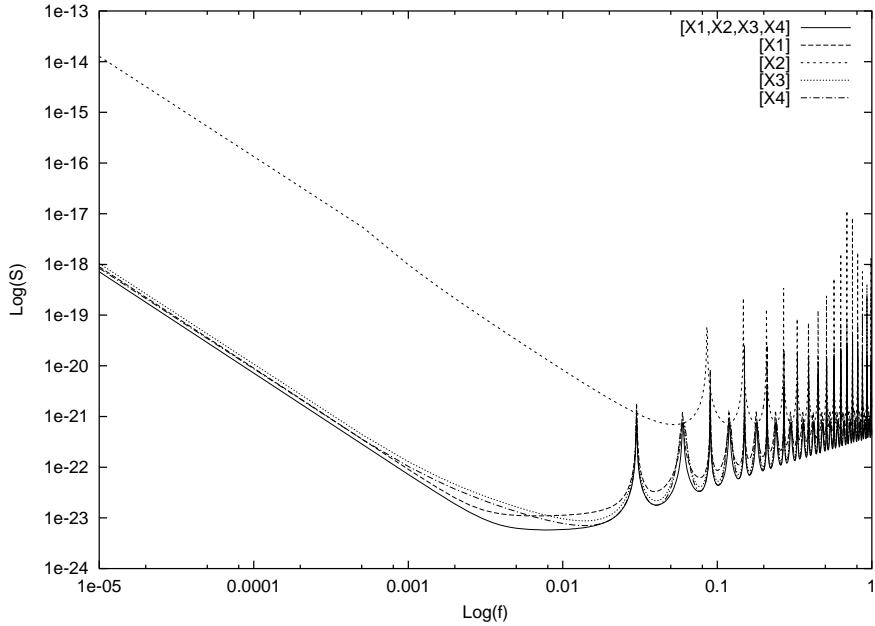}
\end{center}
\caption{$\!\!\!\!\!$(a). Log Log plot of the sensitivity $S$ for the generators $X^{(A)}$ 
as function of $f$ for $\theta=0$, $\phi=0$ over one
year observation period for SNR = 5. The curve [X1,X2,X3,X4] depicts the sensitivity
of the linear  combination of four generators $X^{(A)}$, which gives maximum
SNR.
  } \label{fig5}
\end{figure}
\begin{figure}[t]
\epsfxsize=12cm
\begin{center}
\epsfbox{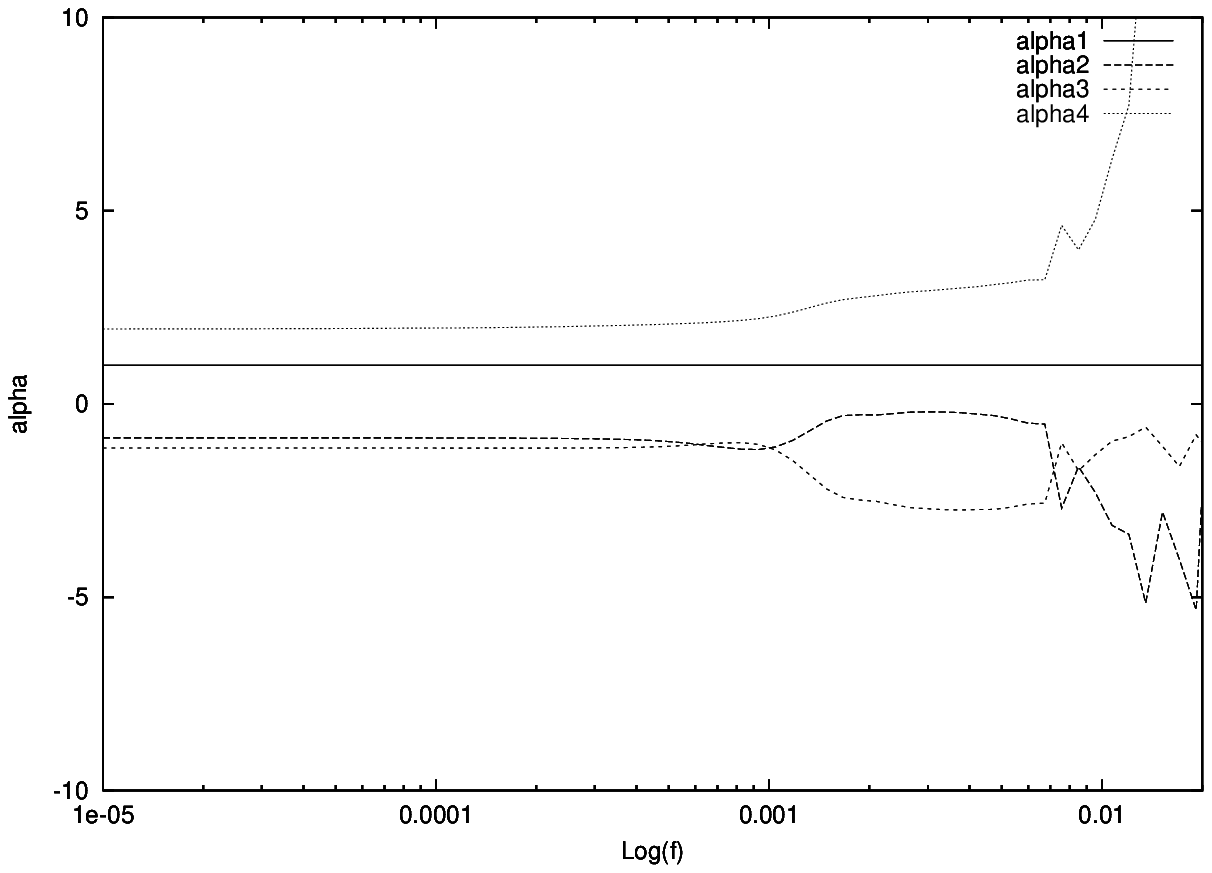}
\end{center}
{FIG. \ref{fig5}(b). Plot of coefficients $\alpha_{(A)}$  which gives the maximum
SNR for linear combinations of all the  four  $X^{(A)}$
as function of $f$ for $\theta=0$ and $\phi=0$.
 }
\end{figure} 
\ \\
\begin{figure}
\epsfxsize=12cm
\begin{center}
\epsfbox{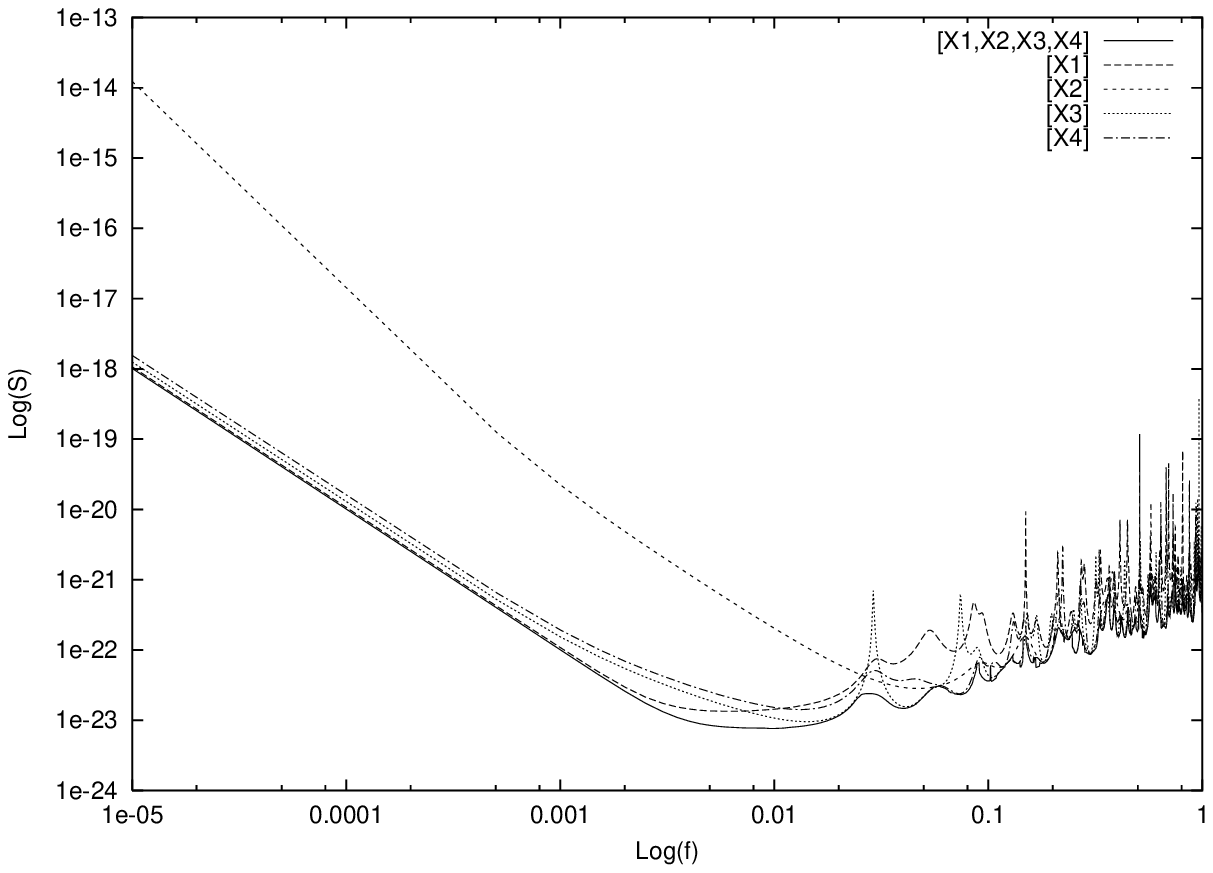}
\end{center}
 \caption{$\!\!\!\!\!$(a). Log Log plot of the sensitivity $S$ for the generators $X^{(A)}$ as function of $f$ for $\theta=\pi/4$, $\phi=\pi/4$
over one  year observation period for SNR = 5.  The curve [X1,X2,X3,X4] depicts 
the sensitivity of the linear  combination of four generators $X^{(A)}$ 
which gives maximum SNR.
   } \label{fig6} 
\end{figure} 
\ \\
\begin{figure}[t]
\epsfxsize=12cm
\begin{center}
\epsfbox{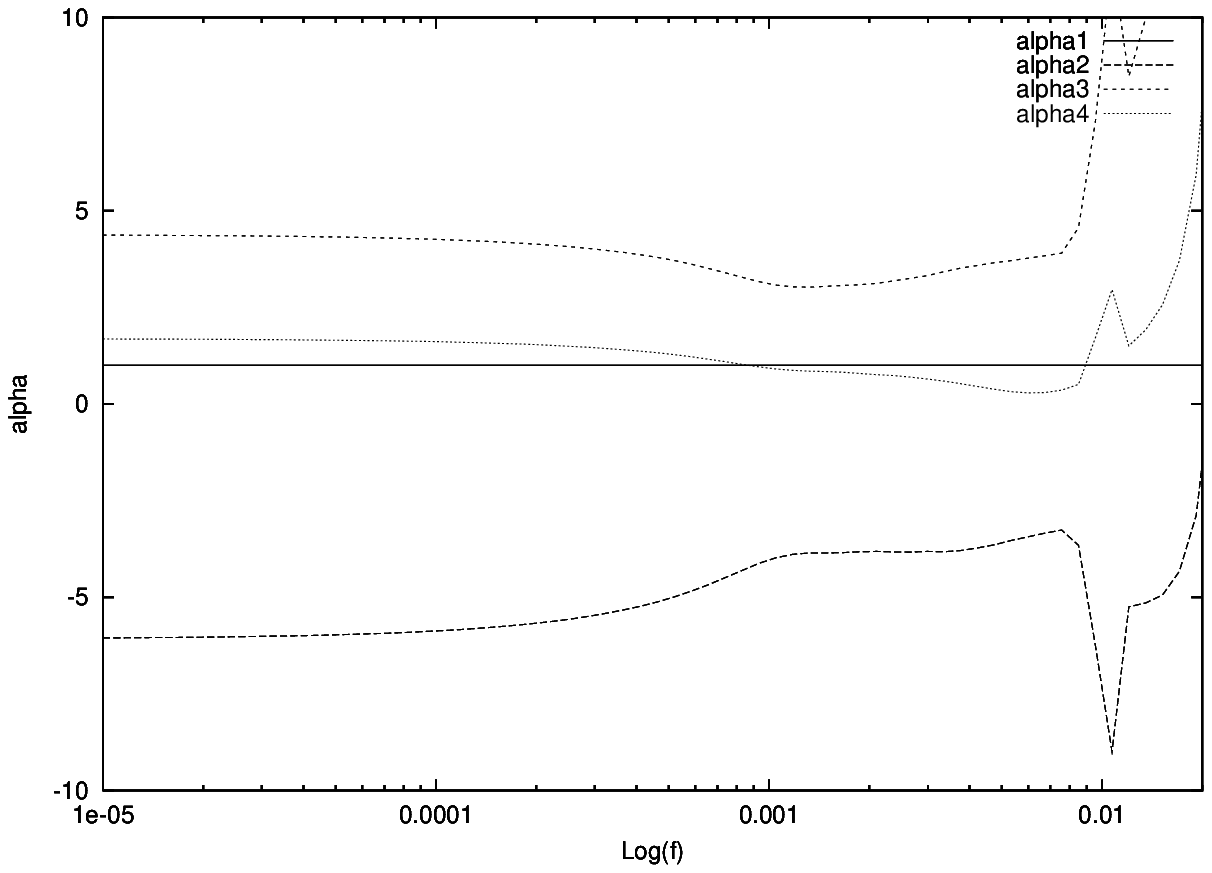}
\end{center}
{FIG. \ref{fig6}(b). Plot of coefficients $\alpha_{(A)}$  which gives the maximum
SNR for linear combinations of all the  four  $X^{(A)}$
as function of $f$ for $\theta=\pi/4$ and $\phi=\pi/4$.
 }
  \end{figure}

This simple case demonstrates that one can use the 
solutions given in the eq. (\ref{gen6}) to get a better SNR. However,
to get full advantage one needs to maximise the SNR over the three $\a$'s.
In order to optimise the SNR given by the general combination  we resort to
numerical methods since  there is no straight forward method for solving the
coupled algebraic equations given by (\ref{max_cond}). We use the  Powell's 
method as given in  \cite{nr} for maximising the 
SNR over the parameters $\alpha_{(A)}$.	
The sensitivity $S$ for the generators $X^{(A)}$ and for the maximal SNR 
combination of $X^{(A)}$'s denoted by [X1,X2,X3,X4] as a function of
frequency $f$ has been plotted in figures \ref{fig5}(a)  and \ref{fig6}(a).
The coresponding values of $\alpha_{(A)}$ are shown in 
 figures \ref{fig5}(b) and \ref{fig6}(b).

\section{Concluding Remarks}
 We have presented in this paper a rigorous and systematic procedure for obtaining data 
combinations which cancel the laser frequency noise based on algebraic geometrical techniques 
and commutative algebra. The data combinations cancelling the laser noise have the structure of 
a module called the module of syzygies. The module is over a ring of polynomials in three 
variables, corresponding to the three time delays along the three arms of the interferometer.
Our formalism can be extended in a straightforward way to include (cancel) the Doppler shifts 
due to the motion of the optical benches. This module provides us with a choice of 
data combinations which are in turn linear combinations of the generators of the module. 
We use this parametrisation to maximise the SNR  over frequency for one class
of GW  sources, namely, those 
that are monochromatic.
We observe that in the plot of sensitivity  verses direction 
angles for the generators $X^{(A)}$, namely figures \ref{fig7} (a)-(d), the sensitivity has 
several peaks. It may be possible to employ this property to
optimally resolve binaries in the confusion noise regime by considering 
suitable data combinations which would be sensitive to specific directions
in the sky. Also here we have investigated monochromatic GW sources. We believe that 
this formalism may also be applied successfully to other type of GW sources eg. 
stochastic GW background.
 
\section{Acknowledgments}

The authors thank IFCPAR (project no. 2204 - 1) under which this work has been carried out. 
The authors are highly  indebted to Himanee Arjunwadkar for pains takingly
explaining the intricacies of the algebraic geometrical ideas required in
this work and also for spending time and  effort for the same.
SVD would like to thank A. Vecchio, S. Phinney, M. Tinto and T. Prince 
for  fruitful discussions.

\newpage

\appendix

\section{generators of the module of syzygies}
\label{syz}

We require the 4-tuple solutions $(p_3, q_1, q_2, q_3)$ to the equation:
\be
(xyz - 1) p_3 + (xz - y) q_1 + x(1 - z^2) q_2 + (1 - x^2) q_3 = 0 \, ,
\label{cnstr}
\ee
where for convenience we have substituted $x = E_1, y = E_2, z = E_3$. $p_3, q_1, q_2, q_3$ 
are polynomials in $x, y, z$ with integral coefficients i.e. in Z[x,y,z].

We now follow the procedure in book by Becker et al. \cite{becker}.

Consider the ideal in $Z[x,y,z]$ (or $Q[x,y,z]$ where $Q$ denotes the field of rational 
numbers), formed by taking linear combinations of the coefficients in 
eq.(\ref{cnstr}) $f_1 = xyz - 1, f_2 = xz - y, f_3 = x(1 - z^2), f_4 = 1 - x^2$. A Groebner 
basis for this ideal is:
\be
{\cal G} = \{g_1 = z^2 - 1, g_2 = y^2 - 1, g_3 = x - yz \}  \, .
\ee
The above Groebner basis is obtained using the function GroebnerBasis in Mathematica.
One can check that both the $f_i, i = 1,2,3,4$ and $g_j, j =1,2,3$ generate the same ideal 
because we can express one generating set in terms of the other and vice-versa:
\be 
f_i = d_{ij} g_j , ~~~~~~~~g_j = c_{ji} f_i \, ,
\ee
where $d$ and $c$ are $4 \times 3$ and $3 \times 4$ polynomial matrices 
respectively, and are given by,
\be
 d = \left(\begin{array}{ccc}
  1 & z^2 & yz \\
  y & 0 & z \\
  -x & 0 & 0 \\
 -1 & - z^2 & -(x + yz) 
\end{array}\right) \, ,
 \hspace{0.3in}
c = \left(\begin{array}{cccc}
  0 & 0 & -x & z^2 - 1 \\
  1 & -y & 0 & 0 \\
  0 & z & 1 & 0 
\end{array}\right) 
\,.
\label{cd}
\ee
The generators of the 4-tuple module are given by the set $A \, \bigcup B^*$ where $A$ and $B^*$ 
are the sets described below:

$A$ is the set of row vectors of the matrix $I - d . c $ where the dot denotes the matrix 
product and I is the identity matrix, $4 \times 4$ in our case. Thus,
\bea
a_1 &=& (1 - z^2, 0, x - yz, 1 - z^2) \, , \no \\
a_2 &=& (0, z(1 - z^2), xy - z, y(1 - z^2)) \, , \no \\
a_3 &=& (0, 0, 1 - x^2, x(z^2 - 1)) \, , \no \\
a_4 &=& (z^2, xz, yz, z^2) \, . 
\eea
We thus first get 4 generators. The additional generators are obtained by computing the 
S-polynomials of the Groebner basis ${\cal G}$.  
The S-polynomial of two polynomials $g_1, g_2$ is obtained by multiplying $g_1$ and 
$g_2$ by suitable terms and then adding, so that the highest terms cancel. For example 
in our case $g_1 = z^2 - 1$ and $g_2 = y^2 - 1$ and the highest terms are $z^2$ for 
$g_1$ and $y^2$ for $g_2$ . Multiply $g_1$ by $y^2$ and $g_2$ by $z^2$ and subtract. 
Thus, the S-polynomial $p_{12}$ of $g_1$ and $g_2$ is:
\be
p_{12} = y^2 g_1 - z^2 g_2 = z^2 - y^2 \, .
\ee
Note that order is defined ($x >> y >> z$) and the $y^2 z^2$ term cancels.
For the Groebner basis of 3 elements we get 3 S-polynomials $p_{12}, p_{13}, p_{23}$.
The $p_{ij}$ must now be rexpressed in terms of the Groebner basis ${\cal G}$.  
This gives a $3 \times 3$ matrix $b$. The final step is to transform to 4-tuples by 
multiplying $b$ by the matrix $c$ to obtain $b^* = b . c$. The row vectors 
$b^*_i , i = 1, 2, 3 $ of $b^*$ form the set $B^*$: 
\bea
b^*_1 &=& (1 - z^2, y(z^2 - 1), x(1 - y^2), (y^2 -1)(z^2 - 1)) \, , \no \\
b^*_2 &=& (0, z(1 - z^2), 1 - z^2 - x(x - yz), (x - yz)(z^2 - 1)) \, , \no \\
b^*_3 &=& (x - yz, z - xy, 1 - y^2, 0) .
\eea
Thus we obtain 3 more generators which gives us a total of 7 generators of the required 
module of syzygies.
 
\section{matrices of conversion between generating sets}
\label{mat}
In this appendix we list the three sets of generators and relations among them.
We first list below $\a, \b, \g, \z$:
\bea
\a &=& (1, z, xz, 1, xy, y) \, , \no \\
\b &=& (xy, 1, x, z, 1, yz) \, , \no \\
\g &=& (y, yz, 1, xz, x, 1) \, , \no \\
\z &=& (x, y, z, x, y, z)  \, . 
\eea

We now express the $a_i$ and $b^*_j$ in terms of $\a, \b, \g, \z$: 
\bea
a_1 &=& \g - z \z \, , \no \\
a_2 &=& \a - z \b \, , \no \\
a_3 &=& - z \a + \b - x \g + xz \z \, , \no \\
a_4 &=& z \z \no \\
b^*_1 &=& - y \a + yz \b + \g - z \z \, , \no \\
b^*_2 &=& (1 - z^2) \b - x \g + xz \z \, , \no \\
b^*_3 &=& \b - y \z \, . 
\eea
Further we also list below $\a, \b, \g, \z$ in terms of $X^{(A)}$:
\bea
\a &=& X^{(3)} \, , \no \\
\b &=& X^{(4)} \, , \no \\
\g &=& - X^{(1)} + z X^{(2)} \, , \no \\
\z &=& X^{(2)} \, . 
\eea
This proves that since the $a_i, b^*_j$ generate the required module, the $\a, \b, \g, \z$ and 
$X^{(A)}, A = 1, 2, 3, 4$ also generate the same module.
\par
The Groebner basis is given in terms of the above generators as follows:
$G^{(1)} = \z, G^{(2)}=  X^{(1)}, G^{(3)} = \b, G^{(4)} = \a$ and
$G^{(5)} = a_3$. 
%
%
%
%

\end{document}